\documentclass[a4paper,11pt]{article}
\usepackage{pos}

\usepackage{lineno}
\usepackage{xspace}
\usepackage{hyperref}
\usepackage{sidecap}

\newcommand{\jpsi} {\ensuremath{{\mathrm J}/\psi}\xspace}
\newcommand{\psiprime} {\ensuremath{\psi(2S)}\xspace}
\newcommand{\pt}   {\ensuremath{p_{\rm T}}\xspace}
\newcommand{\PbPb}         {\mbox{Pb--Pb}\xspace}

\newcommand{\pPb}          {\mbox{p--Pb}\xspace}
\newcommand{\fivenn}       {$\sqrt{s_{\mathrm{NN}}}~=~5.02$~TeV\xspace}

\usepackage{bm}
\usepackage{textcomp}
\usepackage{amsmath}
\newcommand{\textapprox}{\raisebox{0.5ex}{\texttildelow}}

\title{UPC: a powerful tool for \jpsi photoproduction analysis in ALICE}
\ShortTitle{\jpsi photoproduction in UPC}

\author*[a]{Simone Ragoni for the ALICE Collaboration}

\affiliation[a]{University of Birmingham,\\
  B15 2TT, Birmingham, United Kingdom (UK)}

\emailAdd{simone.ragoni@cern.ch}

\abstract{
Ultra-peripheral collisions (UPC) occur when the interacting nuclei or protons have an impact parameter larger than the sum of their radii. They are mediated by  virtual photon exchange. The photoproduction of heavy vector mesons is especially interesting because they couple to the photon.

The ALICE Collaboration has analysed both \pPb and \PbPb UPC at the centre-of-mass energy of $\sqrt{s_{\rm NN}} = 5.02\text{ TeV}$ which correspond to $\gamma$p and $\gamma$-Pb interactions, respectively. This poster reports the exclusive photoproduction of \jpsi off proton and Pb targets,  which shed light on the occurrence of saturation and nuclear shadowing, respectively.

In more detail, the \pPb results measure the growth of the cross section for exclusive production over a wide range in Bjorken-$x$ from \textapprox$10^{-2}$ to \textapprox$10^{-5}$, while \PbPb results demonstrate the presence of nuclear shadowing at high energies and low scales of the order of the mass of the \jpsi.

}

\FullConference{%
  The Eighth Annual Conference on Large Hadron Collider Physics-LHCP2020 \\
  25-30 May, 2020\\
  online}

\begin{document}
\maketitle

\section{A brief introduction to ultra-peripheral collisions}
\begin{SCfigure}[]
    \includegraphics[width = 0.35\textwidth]{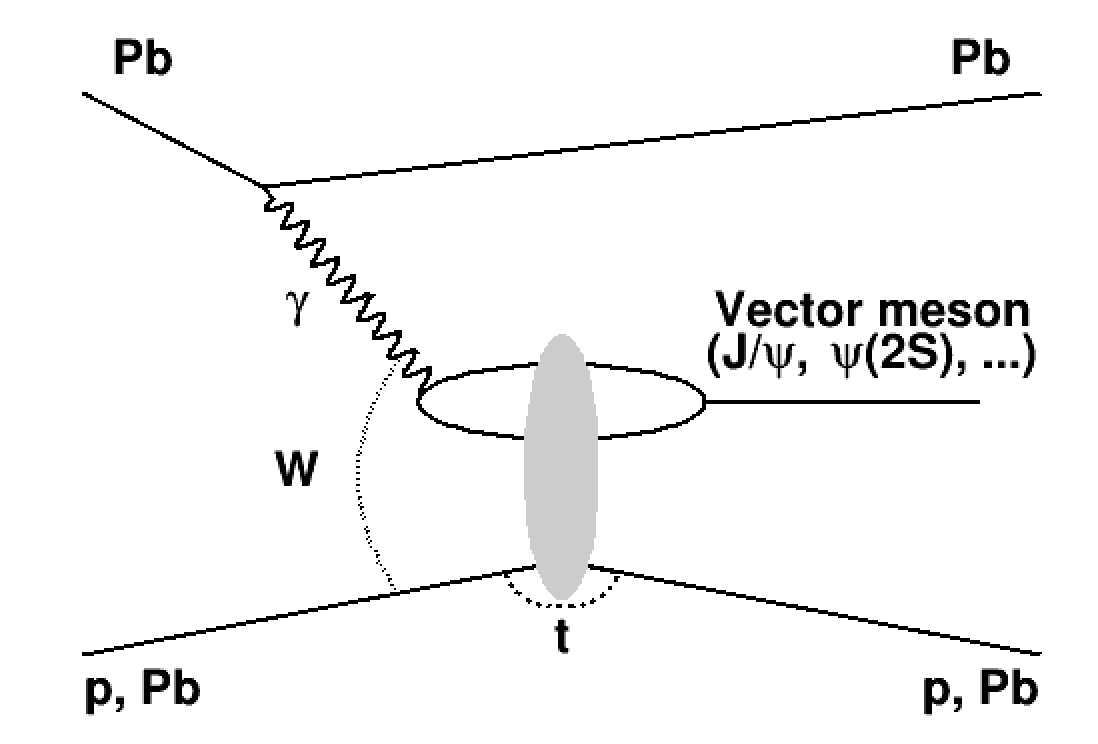}
  \caption{Vector meson photoproduction diagram in UPC. The photon emitted from a Pb ion is interacting with a Pomeron from the target. A vector meson is the end result of such an interaction. The Pomeron is needed to explain the low multiplicity of the process (just the vector meson).}
  \label{fig:diagram}
\end{SCfigure}

Ultra-peripheral collisions (UPC) occur when the interacting ions or protons are beyond the range of the strong interaction. Protons and ions in the beams carry an electromagnetic field and can be viewed as virtual photon sources, thus giving rise to UPC. These photons are quasi-real, and the \textit{Generalised Vector Dominance Model} (GVDM) 
\cite{Schildknecht:2005xr}
predicts vector meson photoproduction with the quantum numbers of the photon, $J^{PC} = 1^{--}$, as shown in Fig.~\ref{fig:diagram}. 

Essentially, the photon interacts with a Pomeron originating from the interaction partner (a colour singlet is required for such a clean process). Pomerons can be regarded as a gluon ladder, so UPC are a sensitive probe of the gluon pdf.
This contribution focuses on \jpsi photoproduction in UPC, which is hence sensitive to Bjorken-$x$ at the level of $10^{-5}$ at \fivenn.

\section{A brief introduction to ALICE}

The ALICE detector has excellent particle identification (PID) capabilities \cite{ALICEpaper}. It can roughly be divided in three parts:
\begin{itemize}
    \item a central barrel, featuring the Inner Tracking System, for vertex determination and PID capabilities at low \pt, the Time Projection Chamber, the Transition Radiation Detector, the Time-Of-Flight system,  two electromagnetic calorimeters and a Cherenkov detector.
    \item the Forward Muon Spectrometer, to detect muons at forward rapidities, which is the main system used for this contribution.
    \item a set of forward detectors, small detectors with acceptances close to beam rapidities, such as the V0, the AD and the ZDC detectors. These detectors are mostly used in UPC for online and offline vetoes.
\end{itemize}

\section{ALICE results }
The ALICE Collaboration has analysed \jpsi photoproduction in \PbPb and \pPb UPC. The focus changes with the selected collision system.

\begin{figure}[ht!]
    \begin{center}
    \includegraphics[width = 0.6\textwidth]{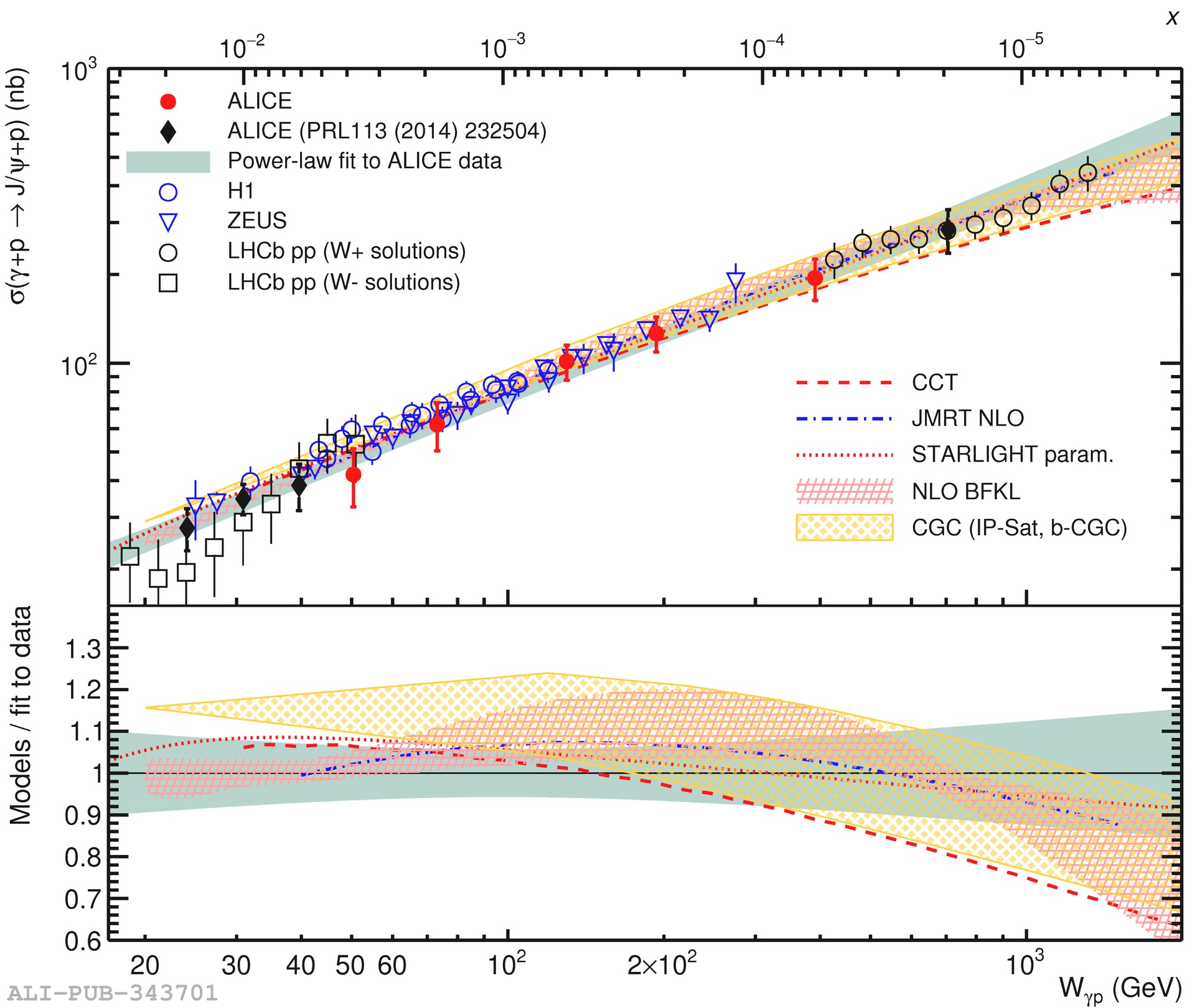}
    \end{center}
  \caption{Exclusive \jpsi photoproduction cross section in \pPb UPC as a function of $W_{\gamma {\rm p}}$, the centre-of-mass energy of the $\gamma {\rm p}$ system. The growth of the cross section can be described by a power-law trend. LHCb solutions in ${\rm p}{\rm p}$ collisions at $\sqrt{s} = 7 \text{ TeV}$ are also shown \cite{LHCbPaper} together with data from HERA \cite{ZEUSpaper, H1paper}. The complete reference list is presented in \cite{Acharya_2019}.
  }
  \label{fig:pPb}
\end{figure}

The process of interest in \pPb is \textit{exclusive vector meson photoproduction}.
The results that the ALICE Collaboration has obtained up to now are shown in Fig.~\ref{fig:pPb}. It shows the growth of the cross section of exclusive \jpsi photoproduction as a function of the centre-of-mass energy of the $\gamma {\rm p}$ system, $W_{\gamma {\rm p}}$. $W_{\gamma {\rm p}}$ is directly related to the Bjorken-$x$ of the process. This is shown in the upper part of the plot. ALICE results have reached a $W_{\gamma {\rm p}}$ of about $700 \text{ GeV}$. This directly translates to a Bjorken-$x$ of about $10^{-5}$ \cite{Acharya_2019}.

The interesting point of Fig.~\ref{fig:pPb} is that the power-law rise of the cross section is interpreted as the growth of the gluon content of the proton. However, it is expected that the rise of the cross section should be tamed by phenomena such as gluon saturation. This would be reflected in the growth of the exclusive photoproduction cross section, as a change from the power-law trend, but such a behaviour is yet to be seen.

The focus in \PbPb UPC events shifts to \textit{coherent} \jpsi photoproduction. As the interaction of the photon will be with another lead nucleus, two types of processes may take place:
\begin{itemize}
    \item \textit{coherent photoproduction}, where the photon interacts coherently with the nucleus;
    \item \textit{incoherent photoproduction}, where the photon interacts with a single nucleon.
\end{itemize}
The two processes have different \pt-distributions due to the different size of the two targets. The average \pt is larger for the incoherent than for the coherent contribution. As a consequence, a simple \pt cut on the dimuon \pt succeeds in selecting a coherently enriched sample.

\begin{figure}[ht!]
    \begin{center}
    \includegraphics[width = 0.6\textwidth]{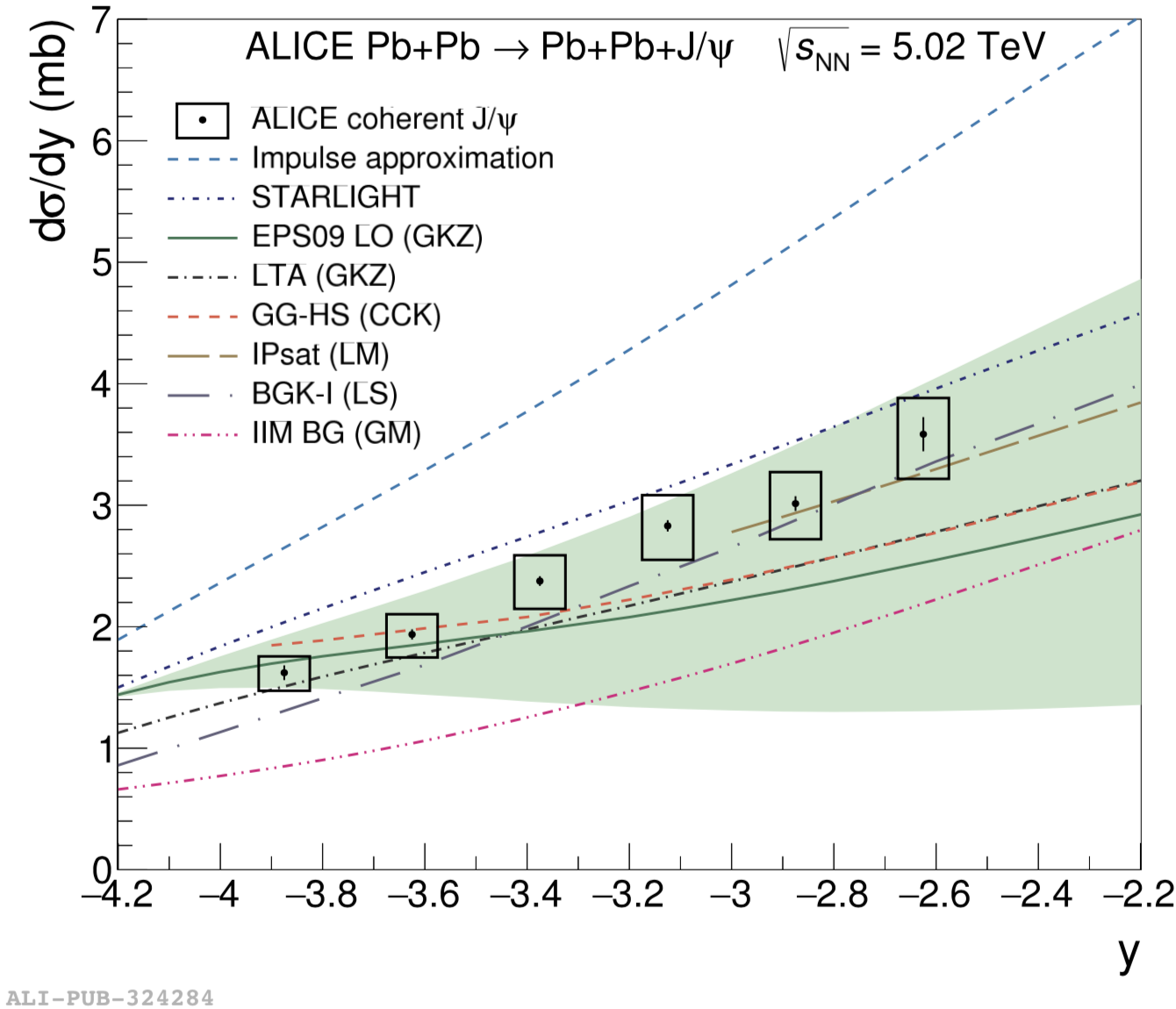}
    \end{center}
  \caption{Coherent \jpsi photoproduction cross section in \PbPb UPC as a function of rapidity \cite{Acharya_2019_2}. The data are compared with the theoretical predictions.}
  \label{fig:shadowing}
\end{figure}

The focus in \PbPb UPC events is to investigate nuclear shadowing, which is a reduction in the photoproduction cross section on the whole nucleus compared to the theoretical prediction when the nucleus is treated as an incoherent sum of nucleons. Fig.~\ref{fig:shadowing} shows the cross section for coherent \jpsi photoproduction as a function of rapidity \cite{Acharya_2019_2}. A comparison with a few available theoretical predictions is also provided. Of particular interest is the comparison against the impulse approximation model. The latter has no implementation of a nuclear shadowing recipe. As such, the data being roughly half of the impulse approximation prediction is compelling evidence of nuclear shadowing effects. Models with moderate shadowing are favoured by the available data.

\end{document}